\documentclass[letterpaper, 10 pt, conference]{ieeeconf}
\IEEEoverridecommandlockouts   % needed to use \thanks
\usepackage{graphicx} % Required for inserting images
\usepackage{url}
\usepackage{svg}
\usepackage{graphics} % for pdf, bitmapped graphics files
\usepackage{epsfig} % for postscri pt graphics files
\usepackage{times} % assumes new font selection scheme installed
\usepackage{amsmath} % assumes amsmath package installed
\usepackage{amsthm}
\usepackage{amssymb}  % assumes amsmath package installed
\usepackage{dsfont}
\usepackage{subcaption}
\usepackage{algorithmic}
\usepackage[ruled,vlined]{algorithm2e}

\usepackage{booktabs}
\usepackage{array}
\usepackage{caption}
\usepackage{float}

\title{Motion Planning with Precedence Specifications via Augmented Graphs of Convex Sets}
\author{Shilin You, Gael Luna, Juned Shaikh, David Gostin, Yu Xiang, Justin Koeln, Tyler Summers
\vspace{-40pt}
\thanks{The authors are with the Department of Mechanical Engineering at University of Texas at Dallas. Email: \texttt{tyler.summers@utdallas.edu}. This work is supported by the United States Air Force Office of Scientific Research under Grant FA9550-23-1-0424, and by the National Science Foundation under Grant ECCS-2047040.}
}
\begin{document}

\maketitle

\begin{abstract}
We present an algorithm for planning trajectories that avoid obstacles and satisfy key-door precedence specifications expressed with a fragment of signal temporal logic. Our method includes a novel exact convex partitioning of the obstacle free space that encodes connectivity among convex free space sets, key sets, and door sets. We then construct an augmented graph of convex sets that exactly encodes the key-door precedence specifications. By solving a shortest path problem in this augmented graph of convex sets, our pipeline provides an exact solution up to a finite parameterization of the trajectory. To illustrate the effectiveness of our approach, we present a method to generate key-door mazes that provide challenging problem instances, and we perform numerical experiments to evaluate the proposed pipeline. Our pipeline is faster by several orders of magnitude than recent state-of-the art methods that use general purpose temporal logic tools. 
% Notes on fast motion planning with temporal logic precedence specifications via graphs of convex sets.
\vspace{-5pt}
\end{abstract}

\section{Introduction}
% context
Motion planning is a fundamental problem in robotics and autonomous systems involving the computation of feasible trajectories for agents navigating through complex environments with various constraints \cite{lavalle2006planning,latombe2012robot}. Many approaches have been developed over several decades, including sampling-based methods, combinatorial graph search, and optimization-based techniques. Due to inherent complexities in motion planning, these approaches must navigate trade-offs between computational efficiency, optimality, and the ability to handle rich geometric, dynamic, and logical constraints.

% overview of most relevant literature
Recent research has developed a promising new framework involving shortest path problems in graphs of convex sets (GCS) \cite{marcucci2024shortest,marcucci2023motion}. This approach elegantly integrates the combinatorial and continuous features whose combination is part of what make motion planning challenging. The combinatorial structure of a graph encodes discrete components, such as region transitions, logical constraints, or contact modes, while each node and edge of the graph is associated with convex sets and functions that describe the continuous spatiotemporal geometry of the problem. This enables powerful tools from convex optimization and mixed-integer optimization that can compute globally optimal or certifiably near-optimal solutions for complex environments and constraints.

\begin{figure*}[t]
  \centering
  \includegraphics[width=0.9\textwidth]{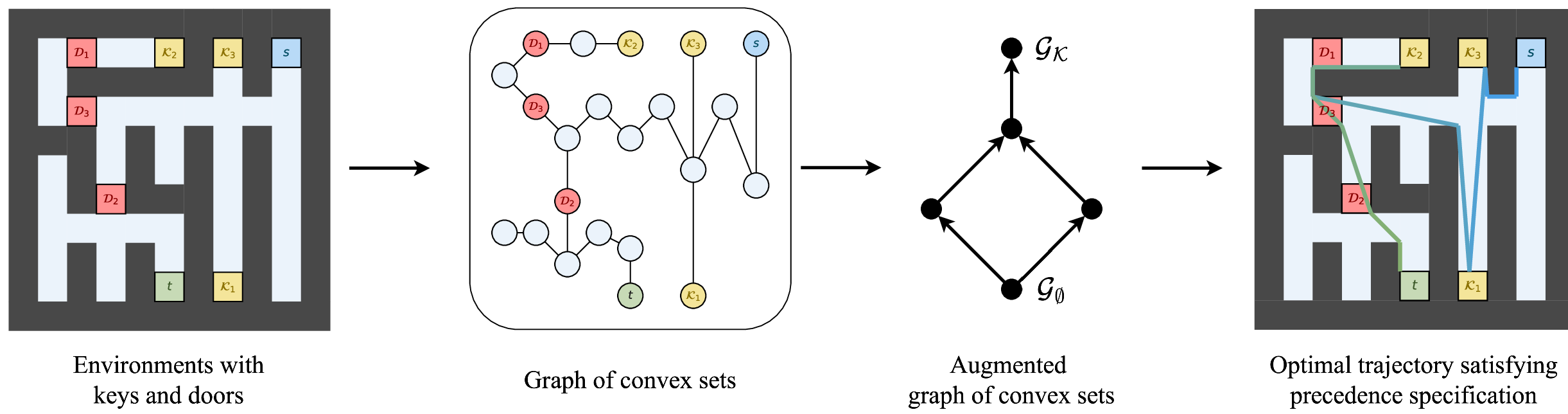}
  \caption{Overview of the proposed approach. The environment is first partitioned and represented as a Graph of Convex Sets (GCS). An augmented GCS is then constructed to encode precedence specifications, and a shortest path problem is solved for the augmented GCS.}
  \label{fig:workflow}
  \vspace{-15pt}
\end{figure*}

We formulate a motion planning problem with precedence specifications and present a solution method based on graphs of convex sets. The GCS framework is particularly well-suited to problems with logical precedence constraints that can be encoded with temporal logic (TL). In particular, we consider a class of environments consisting of obstacles, keys, and doors, where keys can be used to unlock doors to open potentially shorter paths to a terminal goal state. The objective is to find an optimal path from an initial state to a goal state, accounting for keys being able to unlock doors, which may be required to find an optimal path. An overview of the proposed approach is shown in Fig. \ref{fig:workflow}.

Our main contributions are:
\begin{itemize}
    \item We develop an exact convex partitioning of the free space, key, and door sets, which also produces a labeled graph describing connectivity among the free space, key, and door sets (Section III, Algorithm \ref{partition}).
    \item We develop an algorithm to construct an augmented graph of convex sets that exactly encodes key-door precedence logic (Section IV, Algorithm \ref{Algo1}). An exact solution (up to trajectory parameterization) to the motion planning problem with precedence constraints is then obtained by solving a shortest path problem on this augmented graph of convex sets. 
    \item We present a method to generate key-door maze environments and perform numerical experiments to evaluate the performance of the proposed methods. Our results also include instances from \cite{kurtz2023temporal}, where our methods are faster by several orders of magnitude, and a collection of maze environments created with our maze generator. We study the optimality gap and show that on all instances, the solution obtained with our method is within 1\% of the optimal value, and in many cases is globally optimal. Our code will be provided in an open-source repository.
\end{itemize}

Motion planning with general temporal logic constraints using the GCS framework has been recently studied in \cite{kurtz2023temporal}.
Our key insight is that by restricting attention to a specific fragment of temporal logic specifications involving key-door precedence logic and exploiting this structure to build an augmented graph of convex sets that encodes this logic, we achieve several orders of magnitude speed-up over general purpose temporal logic tools (which to our best knowledge represents the state-of-the-art). This ``bottom-up'' approach focuses on specific TL fragments with exploitable structure rather than ``top-down'' approaches that aim to handle arbitrary TL specifications built from a given grammar. Our approach can be adapted to additional temporal logic fragments that can capture a wide variety of practical motion and task planning problems.

\vspace{-6pt}
\subsection*{Related Work}
The main inspirations of the work are \cite{marcucci2023motion,marcucci2024shortest,kurtz2023temporal}. Our problem can also be considered as a type of task and motion planning problem combining geometric and logical components \cite{garrett2021integrated}.

% \vspace{-0.7cm}

\section{Problem Formulation}
Consider a robot operating in a bounded environment $\mathcal{X} \subset \mathbf{R}^d$. The environment contains a set of obstacles $\mathcal{O} = \{\mathcal{O}_i \}_{i=1,...,n_O}$ with $\mathcal{O}_i \subset \mathcal{X}$. The obstacle-free space is denoted by $\mathcal{C} = \mathcal{X} -\mathcal{O}$. The environment also contains a set of doors $\mathcal{D} = \{\mathcal{D}_i \}_{i=1,...,n_D}$, which are obstacles that can be ``unlocked'' and removed from the environment using a corresponding set of keys $\mathcal{K} = \{ \mathcal{K}_i \}_{i=1,...,n_K}$. 
% With slight abuse of notation, we also use $\mathcal{O}, \mathcal{K}$, and $\mathcal{D}$ and their elements to denote the respective obstacle, key, and door sets and their union. 
The environment and all obstacle, key, and door sets are assumed to be polytopes with given half-space representation:
\begin{equation} \nonumber
\begin{aligned}
\mathcal{X} &= \{ x \in \mathbf{R}^d \mid H_{X} x \leq g_{X} \} \\
\mathcal{O}_i &= \{ x \in \mathbf{R}^d \mid H_{O_i} x \leq g_{O_i} \} \\
\mathcal{D}_i &= \{ x \in \mathbf{R}^d \mid H_{D_i} x \leq g_{D_i} \} \\
\mathcal{K}_i &= \{ x \in \mathbf{R}^d \mid H_{K_i} x \leq g_{K_i} \}.
\end{aligned}
\end{equation}
We assume that the number of keys is equal to the number of doors ($n = n_K = n_D$), and that each key is uniquely associated with a door. Without loss of generality, we assume key $i$ unlocks door $i$ for $i=1,...,n_K$. The key-door logic can be expressed by a mapping $\mathbf{D}: 2^{\mathcal{K}}\rightarrow 2^\mathcal{D}$, which specifies the set of doors that are unlocked by a given set of keys. In our setting, this is simply the identity mapping.
% ; we will discuss how our methods can be extended to key-door logic variations.

%[TS: This assumption could be relaxed to allow each key to work with multiple doors, or to have doors that require multiple keys, and so on; this will make the key-door logic and construction of the corresponding augmented graph more complicated.

%DG: An additional type of key/door interaction we could consider, which would be accurate to Zelda mechanics, is that each key can open any door, and a door stays permanently unlocked once opened, but each key can only be used once. This could mean that each door's open/closed logical state partially depends on a state $n_K$ counting the number of keys currently collected.]

Our goal is to find a trajectory $q$ over a time horizon $T$ that solves the following optimization problem:
\begin{equation} \label{optprob}
    \begin{aligned}
        &\underset{q}{\text{minimize}} \quad &&\alpha \int_0^T \| \dot q(t) \|_2 dt + \beta T \\
        &\text{subject to} \quad && q(t) \in \mathcal{C} \cup \mathcal{K} \cup \mathcal{D} \quad \forall t \in [0, T] \\
        & && q(t) \models \phi(\mathcal{K}, \mathcal{D} ) \\
        & && \dot q(t) \in \mathcal{Q} \quad \forall t \in [0, T] \\
        & && q(0) = q_0, \ q(T) = q_T \\
        & && \dot q(0) = \dot q_0, \ \dot q(T) = \dot q_T.
    \end{aligned}
\end{equation}
The objective is a weighted sum of the trajectory length and the time horizon with weights $\alpha$ and $\beta$, respectively. The first constraint requires the trajectory to be in the union of free space sets, key sets, and door sets for all time. The second constraint is a temporal logic specification that enforces the key-door precedence logic to be satisfied. The third constraint requires the velocity to be in the convex set $\mathcal{Q}$ for all time. The remaining constraints specify initial and terminal conditions for the position and velocity.

The optimization problem \eqref{optprob} is infinite-dimensional, so candidate trajectories can be parameterized with piecewise B\'ezier curves defined by a finite set of control points. It is straightforward to include additional objective function terms and constraints that are convex in this parameterization. For example, penalties and constraints on higher-order derivatives promote additional smoothness on the trajectory $q$. Ensuring that the trajectory is differentiable a certain number of times facilitates the design of dynamically feasible trajectories for fully actuated and differentially flat systems.

% \begin{figure}[htbp]
%   \centering
%   \includegraphics[width=0.8\linewidth]{figures/5_key_5_door Curve.png}
%   \caption{An environment with 5 doors and 5 corresponding keys.}
%   \label{fig:complex key door}
% \end{figure}
\vspace{-5pt}
\subsection{Precedence Constraints via Signal Temporal Logic}
We use signal temporal logic (STL) \cite{maler2004monitoring} to encode the key-door precedence specifications. STL is a variation of temporal logic \cite{baier2008principles} that offers a general and powerful framework to specify complex spatiotemporal tasks and constraints. A STL formula can be composed from the following grammar
\begin{equation}
    \varphi ::= \top \mid p \mid \neg \varphi \mid \varphi_{1} \wedge \varphi_{2} \mid \varphi_{1} \vee \varphi_{2} \mid \varphi_{1} \mathcal{U} \varphi_{2} \mid \varphi_{1} \mathcal{R} \varphi_{2},
\end{equation}
by starting from a set of atomic propositions $AP$ with $p\in AP$ and recursively applying boolean operators: $\neg$ (not), $\wedge$ (and), and $\vee$ (or), and temporal operators: $\mathcal{U}$ (until), and $\mathcal{R}$ (release). The until operator $\phi \mathcal{U} \psi$ is satisfied if $\phi$ remains true until $\psi$ becomes true. The release operator $\phi \mathcal{R} \psi$ is satisfied if $\psi$ remains true until and including when $\phi$ becomes true, and if $\phi$ never becomes true, then $\psi$ always remains true; in other words, $\phi$ releases $\psi$. Additional temporal operators can be defined, such as eventually ($\mathcal{F} \varphi := \top \mathcal{U} \varphi $) and always ($G \varphi := \neg \mathcal{F} \neg \varphi$). STL operators and formulas can also be restricted to hold over specific and finite time intervals. 

The key-door precedence logic can be expressed using the release operator. In particular, we utilize a specific fragment of STL of the form:
\[\varphi = (\mathcal{K}_1 \,\mathcal{R}\, \neg \mathcal{D}_1) \wedge(\mathcal{K}_2 \,\mathcal{R}\, \neg \mathcal{D}_2)\wedge \cdots\wedge(\mathcal{K}_n \,\mathcal{R}\, \neg \mathcal{D}_n)\wedge \mathcal{F}(\mathcal{T}),\]
which releases the door constraint when the corresponding key set is entered and requires the terminal condition $\mathcal{T}$ to eventually be reached. Note that this formulation makes picking up the keys optional, unless doing so is required to reach the target. Alternatively, key-door logic that requires keys to be picked up regardless of whether they are needed can be expressed with the until operator:
\[\varphi = (\neg \mathcal{D}_1\,\mathcal{U}\,\mathcal{K}_1) \wedge(\neg \mathcal{D}_2\,\mathcal{U}\,\mathcal{K}_2)\wedge \cdots\wedge(\neg \mathcal{D}_n\,\mathcal{U}\,\mathcal{K}_n)\wedge \mathcal{F}(\mathcal{T}),\]
which prevents passage through the door sets until the corresponding key sets are entered. When the environment and obstacles impose the requirement to pick up the keys in order to reach the target, these specifications are equivalent.

% Good reference: \cite{Belta_Yordanov_Gol_2017}
\vspace{-10pt}
\subsection{Graphs of Convex Sets}
The graph of convex sets (GCS) framework \cite{marcucci2024shortest} consists of a graph $\mathcal{G}(\mathcal{V}, \mathcal{E})$ with vertices $\mathcal{V}$ and edges $\mathcal{E}$. Each vertex $v \in \mathcal{V}$ is associated with a convex set $\mathcal{X}_v$ and a variable $x_v \in \mathcal{X}_v$. For an edge $e=(u,v) \in \mathcal{E}$, its length is given by a nonnegative, convex, and proper function of the corresponding vertex variables $\ell_e (x_u, x_v)$. An edge can optionally also be associated with a constraint $(x_u, x_v) \in \mathcal{X}_e$; this can be equivalently included by allowing the length function to be extended-valued. 

For a given start vertex $s \in \mathcal{V}$ and target vertex $t \in \mathcal{V}$, a path $p$ is a sequence of distinct vertices that connects $s$ to $t$ via an edge subset $\mathcal E_p \subset \mathcal E$. Denoting by $\mathcal{P}$ the set of all paths from $s$ to $t$ in the graph $\mathcal{G}$, the shortest path problem in GCS is stated as
\begin{equation} \label{gcs}
    \begin{aligned}
        &\underset{p, \ x_v}{\text{minimize}} \quad &&\sum_{e=(u,v) \in \mathcal{E}_p} \ell_e(x_u, x_v) \\
        &\text{subject to} \quad && p \in \mathcal{P}, \\
        & && x_v \in \mathcal{X}_v,\quad \forall v \in p, \\
        & && (x_u, x_v) \in \mathcal{X}_e, \quad \forall e \in \mathcal{E}_p.
    \end{aligned}
\end{equation}
This problem seeks to simultaneously find a (discrete) path in the graph from the start vertex to the target vertex and the (continuous) values $x_v$ for each vertex that minimize the total edge length across all edges in the path. Although this problem is computationally difficult in general, a novel and tight mixed-integer formulation was developed in \cite{marcucci2024shortest}. This formulation uses a network flow formulation of the shortest path problem and exploits duality between perspective cones and valid inequality cones to convexify bilinear constraints that arise from the continuous vertex variables. 

Further, a translation from the motion planning problem (without precedence constraints) into a shortest path problem in graphs of convex sets was developed in \cite{marcucci2023motion}. It was also observed that solving a convex relaxation of the mixed-integer formulation of \eqref{gcs} (by simply dropping binary constraints) and utilizing a cheap rounding algorithm allowed certifiably near-optimal or even optimal solutions for many practical motion planning problems. We refer readers to \cite{marcucci2024shortest,marcucci2023motion} for many important details about reformulations.

Following the approaches of \cite{marcucci2023motion,kurtz2023temporal}, we will develop a pipeline to transform our problem \eqref{optprob} into a shortest path problem in a graph of convex sets \cite{marcucci2024shortest}. First, we develop an exact convex partition of the obstacle-free space into a union of convex sets and construct a graph that encodes connectivity among free space, key, and door sets. Then, we propose a method to build an augmented graph of convex sets that encodes the key-door precedence logic. We demonstrate that a shortest path in this augmented graph of convex sets exactly solves our problem \eqref{optprob}.

\vspace{-5pt}

\section{Exact Convex Partitioning of the Free Space, Key, and Door Sets}

This section presents an exact partitioning algorithm to represent the free space $\mathcal{C} =\mathcal{X} -(\mathcal{O} \cup \mathcal{D} \cup \mathcal{K})$ as a union of convex sets of the form $\mathcal{C} = \{\mathcal{C}_i \}_{i=1,...,n_C}$, where the convex free space sets $\mathcal{C}_i \subset \mathcal{X}$ intersect only at boundaries. Based on this partition, we construct a labeled graph $\mathcal{G} (\mathcal{V},\mathcal{E})$, where the vertices $\mathcal{V} =\{\mathcal{V}_{\mathcal{C}} ,\mathcal{V}_{\mathcal{D}} ,\mathcal{V}_{\mathcal{K}}\}$ are labeled as either free space, key, or door sets, and an edge $(u,v) \in \mathcal{E}$ connects vertices $u$ and $v$ if their corresponding convex sets share a facet.
% a continuous path between their corresponding convex sets exists, i.e, the corresponding intersection is not empty. 
% For example, $(u,v) \in \mathcal{E}$ with $u \in \mathcal{V}_\mathcal{C}$ and $v \in \mathcal{V}_\mathcal{K}$ if $\mathcal{C}_{u} \cap \mathcal{K}_{v} \neq \emptyset$.
% \vspace{-15pt}

\subsection{Cell Enumeration for Hyperplane Arrangements}
We first collect all distinct hyperplanes that form the boundaries for the environment $\mathcal{X}$, obstacles $\mathcal{O}$, keys $\mathcal{K}$, and doors $\mathcal{D}$ into a set $\mathcal{H} = \{ (h_i, g_i) \}_{i=1}^{n_h}$, called a \emph{hyperplane arrangement}. Let $\mathcal{S} : \mathcal{X} \rightarrow \{+, - \}^{n_h}$ denote the sign function defined by
$$ \mathcal{S}(x) = \begin{cases} - \quad h_i^\top x \leq g_i \\ + \quad h_i^\top x \geq g_i \end{cases}\quad \text{for} \ i=\{1,...,n_h\}.$$
This function assigns to each point in $\mathcal{X}$ a sign pattern, called a \emph{marking}, that encodes which side of each hyperplane that point lies. For a fixed marking $m$, the set
$$ \mathcal{P}_m = \{ x \in \mathcal{X} \mid \mathcal{S}(x) = m \}$$
is a (convex) polytope, referred to as a \emph{cell} of the hyperplane arrangement. These sets form an exact convex partition of the environment $\mathcal{X}$, including a partition of the free space and the obstacle, key, and door sets.

To create a labeled graph of convex sets that captures all free space, key, and door sets, we must enumerate all nonempty cells of the hyperplane arrangement. The reverse search algorithm \cite{avis1996reverse} can be used to enumerate all cells and their corresponding markings by solving a sequence of linear programs that effectively execute a depth-first search to construct a spanning tree of the cells. More recently, a method that exploits relationships between hybrid zonotopes and ReLU neural networks was proposed in \cite{Exactobsfree2025}. 
% [TODO: Open question: comparison of these approaches for cell enumeration only.]

Let $M(\mathcal{X})$ denote the image of the sign function on the set $\mathcal{X}$, i.e., the set of all markings that define nonempty cells. Although there are $2^{n_h}$ possible markings, only a subset of them define nonempty cells. In fact, for $n_h$ hyperplanes in $d$-dimensional space, Buck demonstrated the bound 
$$|M(\mathbf{R}^d)| \leq \sum_{i=0}^{d} \binom{n_h}{i} = O(n_h^d), $$
with the bound attained when the hyperplanes are in \emph{general position} (no pair of hyperplanes is parallel and no point lies on more than $d$ hyperplanes).
% and $\mathcal{X} = \mathbf{R}^d$.

\subsection{Forming a Graph of Convex Sets from the Cell Enumeration}
By associating a vertex with each cell, we define the vertex set $\mathcal{V}$ for a graph of convex sets. The corresponding edge set can be determined by exploiting an important property of partitions based on hyperplane arrangements and markings. In particular, two cells share a facet if and only if their markings differ in only a single element. Therefore, for two cells $\mathcal{P}_{m_1}$ and $\mathcal{P}_{m_2}$, there is an edge connecting them if $\| m_1 - m_2 \| = 2$. Performing this comparison for each pair of cells generates the edge set $\mathcal{E}$.

It is easy to distinguish and label cells and vertices associated with key and door sets using the markings. For example, let $H \subset \mathcal{H}$ be the subset of hyperplanes that define the boundary of any key or door. Then set of markings with a $-$ in the corresponding entries, namely
$$M_H =  \{ m \in M(\mathcal{X}) \mid m_i = - \quad     \forall i \in H \} $$
correspond to cells whose union reconstructs that key or door set. In the graph, we merge all vertices associated with this set of markings. Performing this operation for each key and door generates the labeled vertex sets $\mathcal{V}_\mathcal{K}$ and $\mathcal{V}_\mathcal{D}$. Similarly, all cells and corresponding nodes associated with obstacle cells can be identified and removed from the graph. The remaining vertices correspond to free space and define $\mathcal{V}_\mathcal{C}$.

Thus, we have constructed a labeled graph of convex sets $\mathcal{G}(\mathcal{V},\mathcal{E})$ with $\mathcal{V} = \{\mathcal{V}_\mathcal{C}, \mathcal{V}_\mathcal{K}, \mathcal{V}_\mathcal{D} \}$, where the polytopes corresponding to $\mathcal{V}_\mathcal{C}$ form an exact convex partition of the free space, and the edge set encodes connectivity among free space, key, and door sets that share a facet.

\subsection{Merging Free Space Cells to Reduce Graph Size}
Although the graph from the previous section can immediately be used within the GCS motion planning framework, it may contain a large number of free space nodes, depending on the number of obstacles, keys, and doors, and especially the dimension $d$. The optimal complexity reduction problem is to find a \emph{minimal} partition of the free space. This problem was studied in \cite{geyer2008optimal}, where a branch-and-bound algorithm was developed to obtain a partition with a minimal number of polytopes by merging cells that preserve convexity. Alternatively, greedy algorithms can be used to quickly reduce the number of polytopes without necessarily obtaining a minimal partition \cite{geyer2008optimal,Exactobsfree2025}. Our algorithm for exact convex partitioning and labeled GCS construction is summarized in Algorithm \ref{partition}.
    \begin{algorithm}[ht] 
        \caption{Exact Convex Partition \& Labeled GCS Construction}
        \begin{algorithmic}[1] \label{partition}
            \renewcommand{\algorithmicrequire}{\textbf{Input:}}
            \renewcommand{\algorithmicensure}{\textbf{Output:}}
            \REQUIRE Environment $\mathcal{X}$, obstacles $\mathcal{O} = \{\mathcal{O}_i \}_{i=1,...,n_O}$, keys, $\mathcal{K} = \{\mathcal{O}_i \}_{i=1,...,n}$, doors $\mathcal{D} = \{\mathcal{D}_i \}_{i=1,...,n}$ all given in half-space representation
            \ENSURE Labeled graph of convex sets $\mathcal{G}( \mathcal{V}, \mathcal{E})$ with $\mathcal{V} = \{\mathcal{V}_\mathcal{C}, \mathcal{V}_\mathcal{K}, \mathcal{V}_\mathcal{D} \}$ and exact convex partition $\mathcal{X} - \mathcal{O} \cup \mathcal{K} \cup \mathcal{D} = \{\mathcal{C}_i \}_{i=1}^{n_C}$
            \STATE Form hyperplane arrangement from all boundaries
            \STATE Enumerate cells in environment and generate markings $M(\mathcal{X})$ (via reverse search or hybrid zonotopes)
            \STATE Define vertex for each cell to form initial vertex set $\mathcal{V}$
            \STATE Define edge set $\mathcal{E}$ between vertices whose markings differ in exactly one entry
            \FOR {$\mathcal{K}_i \in \mathcal{K}$, $\mathcal{D}_i \in \mathcal{D}$}
            \STATE Merge and label corresponding cells and vertices
            \ENDFOR
            \FOR {$\mathcal{O}_i \in \mathcal{O}$}
            \STATE Merge and discard corresponding cells and vertices
            \ENDFOR
            \STATE Merge free space nodes that preserve convexity (via optimal complexity reduction or greedy algorithm)
            \RETURN $\mathcal{G}( \mathcal{V}, \mathcal{E})$ with $\mathcal{V} = \{\mathcal{V}_\mathcal{C}, \mathcal{V}_\mathcal{K}, \mathcal{V}_\mathcal{D} \}$
        \end{algorithmic}
    \end{algorithm}

% ===========================================
% To compute the exact convex partitioning of the free space environment and generate the labeled graph, the algorithm presented in \cite{Exactobsfree2025} is adapted and modified to include the key and door sets as part of the partition.
% %although non-convex sets can either be decomposed into multiple convex sets or over-approximated. 
% The modified problem can be posed as finding the exact convex partitioning of the set $\mathcal{C} =\mathcal{X} -(\mathcal{O} \cup \mathcal{D} \cup \mathcal{K})$. The key concept for finding the exact convex partitioning of the free space environment in \cite{Exactobsfree2025} leverages the fact that for the hyperplane arrangement $\mathcal{A} \subset \mathbb{R}^{d}$ of $n_{h}$ hyperplanes which represent the disticnt boundaries of all the cells in $\mathcal{O}_{lb}$ and the bounded environment $\mathcal{X}$, a Mixed Integer Linear Program (MILP) can be formulated and solved to identify $\overline{n}_{C}$ freespace cells efficiently, where $\overline{n}_{C}>n_{C}$. The solution of the MILP can be used to construct the free space and the labeled set components of the labeled graph $\mathcal{G}$, which are then combined together into a single labeled graph through adjacency matrix merging and augmentation. Further merging of these cells is then performed using the algorithm used in \cite{Exactobsfree2025} to reduce $\overline{n}_{C}$ to $n_{C}$. The details of the MILP formulation and the adjacency matrix augmentation are presented in the Appendix. 

\section{Construction of an Augmented Graph of Convex Sets Encoding Precedence Logic}
In this section, we describe how to construct an augmented graph of convex sets that encodes the key-door precedence constraints based on the labeled GCS $\mathcal{G}(\mathcal{V}, \mathcal{E})$ from Section III. The augmented GCS consists of copies of $\mathcal{G}(\mathcal{V}, \mathcal{E})$, with different edge sets according to the current set of collected keys. By construction, every path connecting the starting node and a target node copy in the augmented GCS satisfies the key-door precedence constraints. 
% We also describe the construction of similar augmented GCS for variations of the key-door precedence logic.

\subsection{Constructing the Augmented GCS}
The augmented GCS is built recursively from the labeled graph $\mathcal{G}(\mathcal{V}, \mathcal{E})$ with $\mathcal{V} = \{\mathcal{V}_{\mathcal{C}} ,\mathcal{V}_{\mathcal{D}} ,\mathcal{V}_{\mathcal{K}}\}$ based on a start node $s \in \mathcal{V}_{\mathcal{C}}$ whose corresponding set contains the initial condition. The augmented GCS consists of $n_K + 1$ layers, where each layer represents reachable key subsets of a certain cardinality.

\begin{figure}[htbp]
  \centering
  \includegraphics[width=0.7\linewidth]{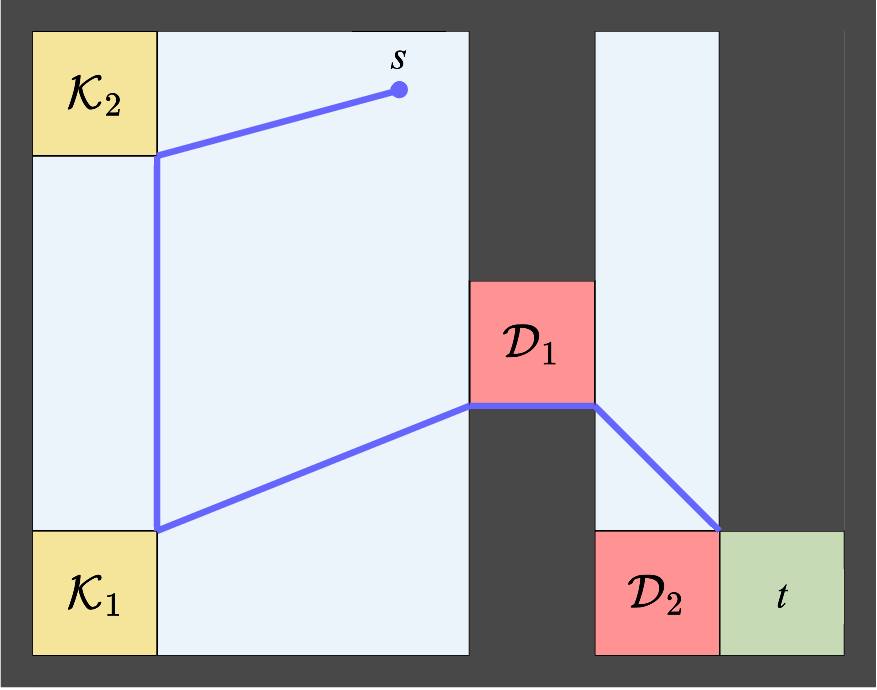}
  \caption{A key-door environment with 2 keys and 2 doors. The blue line shows the optimal solution.}
  \label{fig:simple key door}
\end{figure}

\begin{figure}[htbp]
  \centering
  \includegraphics[width=0.9\linewidth]{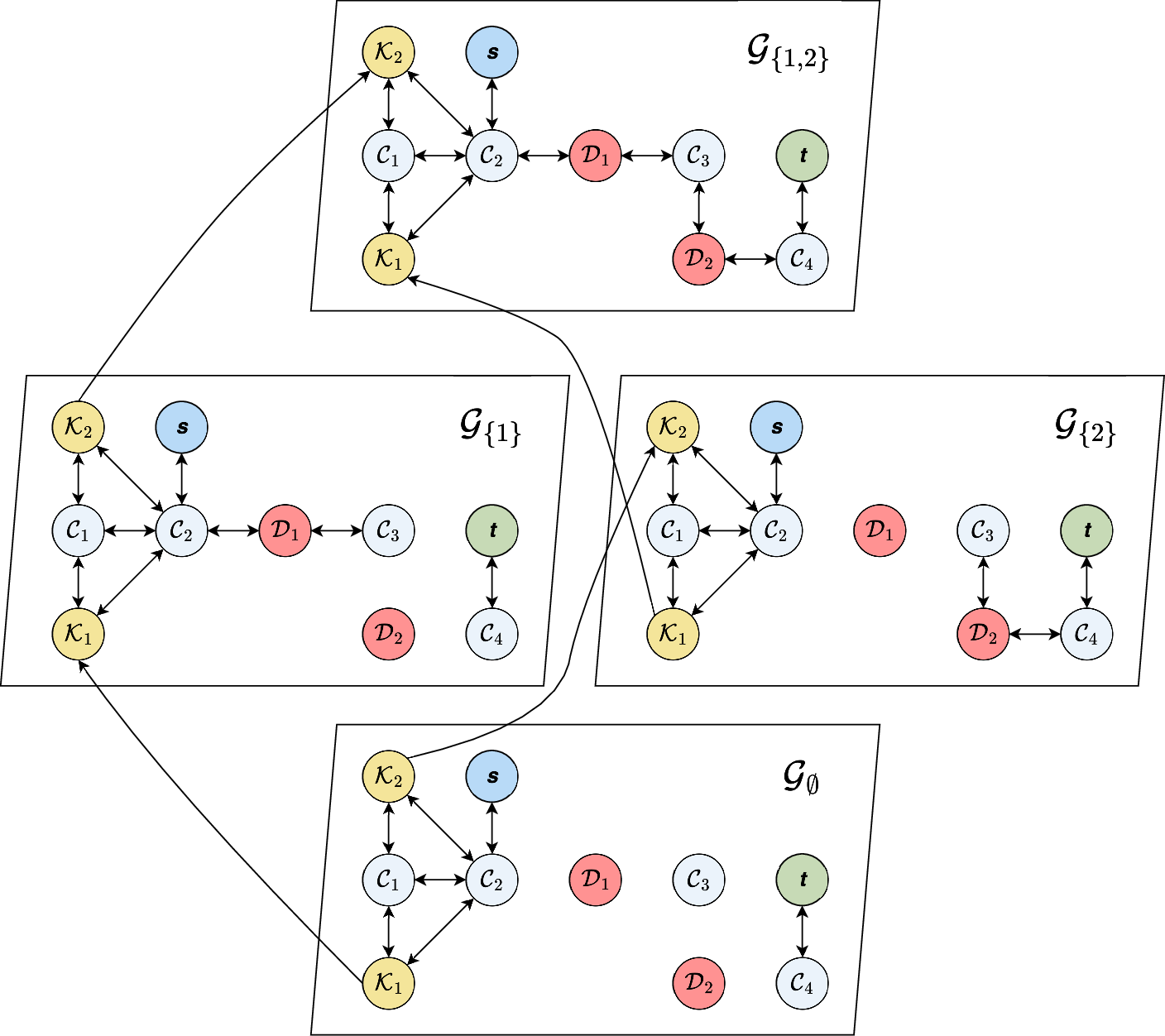}
  \caption{The augmented GCS for a simple 2-key environment. The start point is located in free space node $\mathcal{C}_2$ and the target node is located in $\mathcal{C}_4$. 
  }
  \label{fig:augGCS}
  \vspace{-5pt}
\end{figure}

A simple 2-key environment is shown in Fig. \ref{fig:simple key door}, and the associated augmented GCS is shown in Fig. \ref{fig:augGCS}. We will refer to this as a running example throughout our description.

\textbf{The Base Layer.} The base layer of the augmented GCS corresponds to the empty key set $\emptyset \in 2^\mathcal{K}$. We define the base graph $\mathcal{G}_\emptyset(\mathcal{V}_\emptyset, \mathcal{E}_\emptyset)$. The vertex set $\mathcal{V}_\emptyset$ consists of a copy of all vertices in 
$\mathcal{V}$. The edge set is $$\mathcal{E}_\emptyset = \mathcal{E} - \{ (u, v) \in \mathcal{E} \mid u \in \mathcal{V}_\mathcal{D} \ \text{or} \ v \in \mathcal{V}_\mathcal{D} \},$$
corresponding to the original edge set with all edges adjacent to door nodes removed. The base layer for the simple 2-key environment is shown at the bottom of Fig. \ref{fig:augGCS}.

\textbf{Layer 1.} The next layer of the augmented GCS corresponds to 1-element key sets that are reachable from the start node in the base graph $\mathcal{G}_\emptyset$. Let $S_1 \subset 2^\mathcal{K}$ denote the set of 1-element key sets, and let $k_1 = |S_1|$. The reachable keys can be found using breadth- or depth-first search on the base graph $\mathcal{G}_\emptyset$. For each key subset $S \in S_1$, we create a subgraph $\mathcal{G}_S(\mathcal{V}_S, \mathcal{E}_S)$. The vertex subset $\mathcal{V}_S$ has a copy of all vertices in $\mathcal{V}$. The edge set is $$\mathcal{E}_S = \mathcal{E}_\emptyset \cup \{ (u, v) \in \mathcal{E} \mid u \in \mathbf{D}(S) \ \text{or} \ v \in \mathbf{D}(S) \},$$
which reinserts edges adjacent to the door corresponding to the key in $S$, denoted by $\mathbf{D}(S)$.
We also add a directed edge to each reachable key node copy in $\mathcal{V}_S$ from the corresponding key node copy in $V_\emptyset$. Since the key sets in configuration space associated with each key node copy are identical, these directed edges incur zero cost. (Alternatively, the corresponding key nodes in $\mathcal{V}_S$ and $\mathcal{V}_\emptyset$ can be \emph{merged} into a single node.)

In our running 2-key example, both keys are reachable from the start node, so $S_1 = \{ \{1\}, \{2\} \}$ and $k_1 = 2$. The subgraphs $G_{\{1\}}$ and $G_{\{2\}}$ are shown in the middle of Fig. \ref{fig:augGCS}, along with the directed edges from key nodes in $G_\emptyset$.

\textbf{Layer 2.} The next layer corresponds to 2-element key sets that are reachable from each of the 1-element key sets in Layer 1. Let $S_2 \subset 2^\mathcal{K}$ denote the set of 2-element key sets, and let $k_2 = |S_2|$. The 2-element reachable keys sets can be found  breadth- or depth-first search within each subgraph from layer 1. As in layer 1, for each key subset $S \in S_2$, we create a subgraph $\mathcal{G}_S(\mathcal{V}_S, \mathcal{E}_S)$ with vertex set $\mathcal{V}_S$ featuring a copy of each vertex in $\mathcal{V}$ and edge set given by $$\mathcal{E}_S = \mathcal{E}_\emptyset \cup \{ (u, v) \in \mathcal{E} \mid u \in \mathbf{D}(S) \ \text{or} \ v \in \mathbf{D}(S) \},$$
which reinserts edges adjacent to the doors corresponding to the keys in $S$, denoted by $\mathbf{D}(S)$.
For each subgraph $\mathcal{G}_S(\mathcal{V}_S, \mathcal{E}_S)$ and for each key node $v \in \mathcal{V}_S$, we add directed edges from the copy of key in subgraph $\mathcal{G}_{S - \{v\} }$ in the previous layer.

In our running 2-key example, the only 2-element key set is the full key set, so $S_2 = \{ \{1, 2\} \}$ and $k_2 = 1$. The subgraph $G_{\{1, 2\}}$ is shown at the top of Fig. \ref{fig:augGCS}.

\textbf{Layer $\ell$.} In general, layer $\ell$ consists of subgraphs corresponding to all $\ell$-element key sets that are possible to reach from the starting node. Let $S_\ell$ denote the set of $\ell$-element key sets. The reachable key sets are found by applying breadth- or depth-first search from each subgraph at the previous layer. For each key subset $S \in S_\ell$, we create a subgraph $\mathcal{G}_S(\mathcal{V}_S, \mathcal{E}_S)$ with vertex set $\mathcal{V}_S$ featuring a copy of each vertex in $\mathcal{V}$ and edge set given by $$\mathcal{E}_S = \mathcal{E}_\emptyset \cup \{ (u, v) \in \mathcal{E} \mid u \in \mathbf{D}(S) \ \text{or} \ v \in \mathbf{D}(S) \},$$
which reinserts edges adjacent to the doors corresponding to the keys in $S$. For each subgraph $\mathcal{G}_S(\mathcal{V}_S, \mathcal{E}_S)$ and for each key node $v \in \mathcal{V}_S$, we add directed edges from the copy of key in subgraph $\mathcal{G}_{S - \{v\} }$ in the previous layer.

\textbf{Layer $n_K$.} The final layer corresponds to the full key set $\mathcal{K} \in 2^\mathcal{K}$ and consists of a single subgraph. The final graph $\mathcal{G}_\mathcal{K}(\mathcal{V}_\mathcal{K}, \mathcal{E}_\mathcal{K})$ again has a vertex set $\mathcal{V}_\mathcal{K}$ with a copy of each vertex in $\mathcal{V}$ and has edge set $\mathcal{E}_\mathcal{K} = \mathcal{E}$, the same edge set as $\mathcal{G}$. For each key node $v \in \mathcal{V}_\mathcal{K}$, we add a directed edge from the corresponding key node in the graph $\mathcal{G}_{\mathcal{K} - \{v\} }$.

Finally, given a target node $t \in \mathcal{V}_{\mathcal{C}}$ whose corresponding set contains the terminal condition, we merge all copies of the target node throughout all layers and subgraphs into a single node that serves as the target node in the augmented GCS. The algorithm for constructing the augmented graph is summarized in Algorithm \ref{Algo1}.

For each subgraph in the augmented graph, we associate the free space partition sets, key sets, and door sets with the corresponding copies of the nodes to obtain the augmented GCS. The machinery of the GCS motion planning framework \cite{marcucci2023motion,marcucci2024shortest} can now be applied to compute a shortest path from the start node to the (merged) target node in the augmented GCS. We now discuss the properties of these reformulations for the augmented GCS.

\begin{algorithm}[ht] 
    \caption{Augmented GCS Construction}
    \begin{algorithmic}[1] \label{Algo1}
        \renewcommand{\algorithmicrequire}{\textbf{Input:}}
        \renewcommand{\algorithmicensure}{\textbf{Output:}}
        \REQUIRE Labeled graph $\mathcal{G}(\mathcal{V}, \mathcal{E})$, with $\mathcal{V} = \{\mathcal{V}_\mathcal{C}, \mathcal{V}_\mathcal{D}, \mathcal{V}_\mathcal{K} \}$, start node $s \in \mathcal{V}_\mathcal{C}$, target node $t \in \mathcal{V}_\mathcal{C}$
        \ENSURE Augmented graph $\hat{\mathcal{G}}( \hat{\mathcal{V}}, \hat{\mathcal{E}})$
        \STATE Create subgraph $\mathcal{G}_\emptyset(\mathcal{V}_\emptyset, \mathcal{E}_\emptyset)$, where $\mathcal{V}_\emptyset := \mathcal{V}$, $\mathcal{E}_\emptyset := \mathcal{E} - \{ (u, v) \in \mathcal{E} \mid u \in \mathcal{V}_\mathcal{D} \ \text{or} \ v \in \mathcal{V}_\mathcal{D} \}$ 
        \FOR {$\ell=1$ to $n_K$}
        \STATE Find all reachable $\ell$-element key sets $S_\ell \subset 2^\mathcal{K}$ from start node copy in all subgraphs with $(\ell-1)$-element key sets
        \FOR {$S \in S_\ell$}
        \STATE Create subgraph $\mathcal{G}_S(\mathcal{V}_S, \mathcal{E}_S)$, where $\mathcal{V}_S := \mathcal{V}$, $\mathcal{E}_S := \mathcal{E}_\emptyset \cup \{ (u, v) \in \mathcal{E} \mid u \in \mathbf{D}(S) \ \text{or} \ v \in \mathbf{D}(S) \} $
        \FOR {$v \in S$}
        \STATE Add directed edge $\mathcal{E}_S \leftarrow \mathcal{E}_S \cup \{ (u, v) \} $, where $u$ is the copy of $v$ in  $\mathcal{V}_{S - \{ v \} }$
        \ENDFOR
        \ENDFOR
        \ENDFOR
        \STATE Merge copies of target node in all layers \& subgraphs
        \RETURN $\hat{\mathcal{G}}( \hat{\mathcal{V}}, \hat{\mathcal{E}})$, where $ \hat{\mathcal{V}} = \cup_{S \in \{ S_\ell \}_{\ell = 0}^{n_K}} \mathcal{V}_S $, $ \hat{\mathcal{E}} = \cup_{S \in \{ S_\ell \}_{\ell = 0}^{n_K}} \mathcal{E}_S $ 
    \end{algorithmic}
\end{algorithm}

\subsubsection{Properties of the augmented GCS}
By construction, every path in the augmented GCS from the start node to the (merged) target node satisfies the key-door precedence logic. Therefore, a shortest path in the augmented GCS corresponds to an optimal solution of the original problem \eqref{optprob}. The mixed-integer convex reformulation of the augmented GCS based on \cite{marcucci2024shortest,marcucci2023motion} thus exactly solves \eqref{optprob}, up to a finite parameterization of the trajectory $q$ (e.g., using B\'ezier curves). It is guaranteed to find a path if one exists (in the absence of a bound on the horizon $T$), and certifies infeasibility if a path does not exist. Infeasibility is certified if the (merged) target node is disconnected from the start node in the augmented graph. 

A convex relaxation of the exact mixed-integer formulation, obtained by simply dropping binary constraints, along with an inexpensive rounding scheme, can be used to obtain approximate solutions to \eqref{optprob}. It was observed in \cite{marcucci2023motion} that in many practical motion planning problems (without precedence specifications), this relaxation is often tight and provides certifiably near-optimal solutions. We will study the tightness of this relaxation for problems with precedence specifications in our numerical experiments section.

Further, even the convex relaxation can be challenging to solve in practice. It may be computationally expensive to find an exact convex partition from Section III with the minimum number of free space nodes. Also, the infinite-dimensional trajectory $q$ is discretized with B\'ezier curves of finite order.

% [TODO: Discussion about comparison between augmented GCS and the Deterministic Finite Automaton (DFA) from the Kuntz + Lin paper.]

% augmented GCS is similar to an unfold Deterministic Finite Automaton (DFA) but considered physical constraints in the environment. (...)

\subsubsection{Size of the augmented GCS}
The number of subgraphs in the augmented GCS depends on the number of reachable key sets. Let us obtain an upper bound on the number of subgraphs in the augmented GCS. For an environment with $n_K$ keys and associated doors, recall that $S_k \in 2^\mathcal{K}$ denotes the set of reachable $k$-element key subsets, each of which corresponds to a subgraph in the augmented GCS. The total number of subgraphs satisfies
\begin{equation}
\sum_{k=0}^{n_K} |S_k| \leq \sum_{k=0}^{n_K} \binom{n_k}{k} = 2^{n_K}.
\end{equation}
To see this, at the $k$th layer in the augmented graph, there are at most $\binom{n_k}{k}$ possible reachable key subsets. The bound is obtained when every key is reachable from the starting node. Thus, in the worst case, the total number of subgraphs is exponential in the number of keys.

\begin{figure}[htbp] \label{keysubgraphs}
  \centering
  \includegraphics[width=0.85\linewidth]{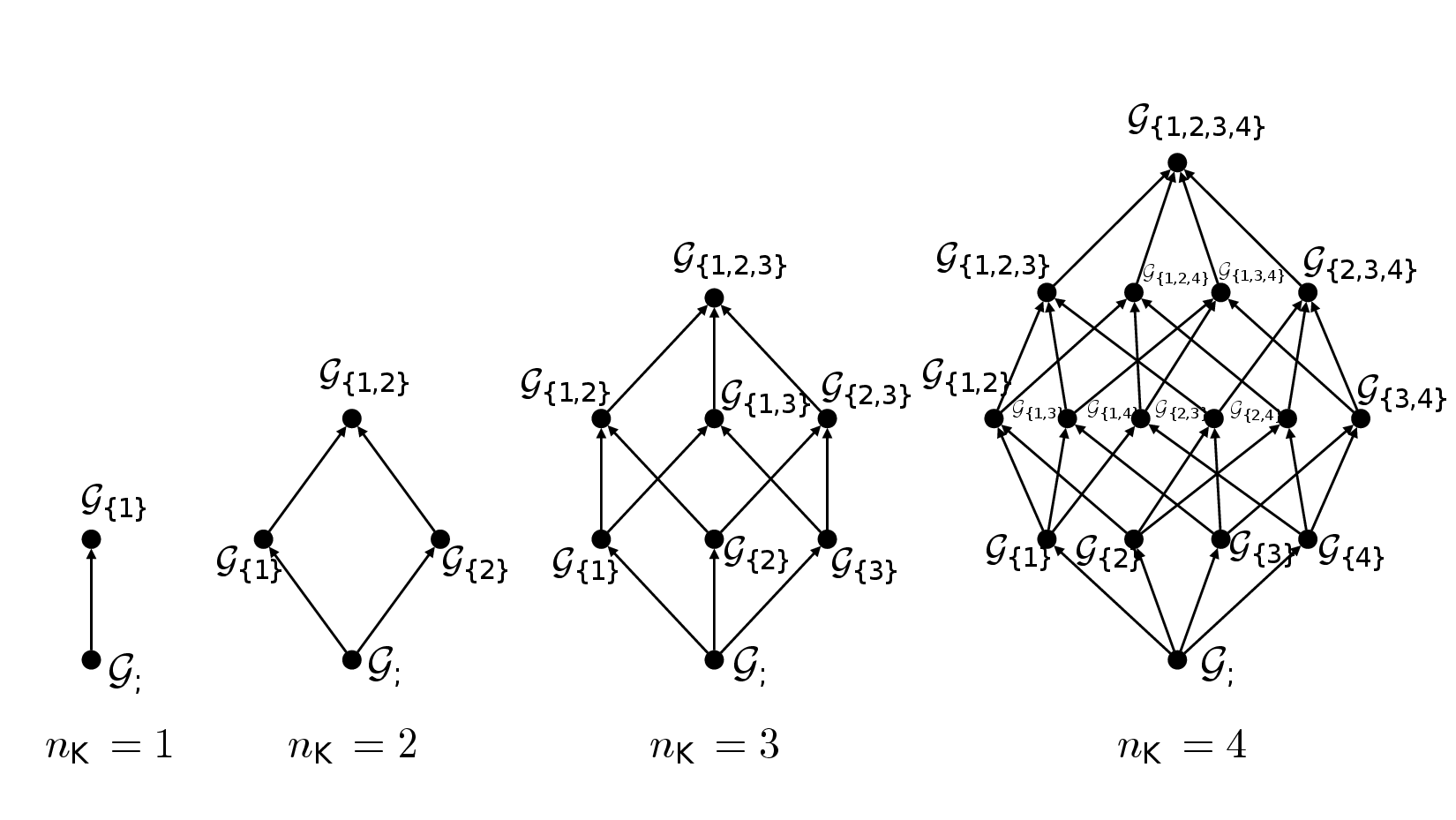}
  \caption{Figure illustrating the size of the augmented GCS through graphs of subgraphs up to $n_K = 4$ keys.}
  \label{fig:subgraphs}
  \vspace{-15pt}
\end{figure}

For a lower bound on the number of subgraphs in the augmented graph, note that there are degenerate cases where some keys are not reachable from the start node. For example, if none of the keys are reachable from the start node, there is only a single copy of the labeled graph in the augmented graph, and the problem reduces to a shortest path problem on the original graph. When only a single additional key is reachable at each layer (i.e., the $S_k$ each contain only one element, meaning that the keys must be obtained in a specific sequence), then there are only $n_K + 1$ subgraphs in the augmented graph. In many practical problems, there are a small number of keys, and the augmented graph has limited width. Fig. \ref{fig:subgraphs} shows the graphs of subgraphs for up to 4 keys when all key subsets are reachable. 

\vspace{-10pt}

\vspace{-5pt}
\section{Numerical Experiments and Key-Door Maze Generator}

% \begin{table*}
% \centering
% \caption{Numerical Experiments with Key-Door Mazes}
% \begin{tabular}{>{\centering\arraybackslash}p{0.03\textwidth}
%                 >{\centering\arraybackslash}p{0.05\textwidth}
%                 >{\centering\arraybackslash}p{0.05\textwidth}
%                 >{\centering\arraybackslash}p{0.08\textwidth}
%                 >{\centering\arraybackslash}p{0.08\textwidth}
%                 >{\centering\arraybackslash}p{0.15\textwidth}
%                 >{\centering\arraybackslash}p{0.10\textwidth}
%                 >{\centering\arraybackslash}p{0.14\textwidth}}
% \toprule
% ID & $|\mathcal{V}|$ & $|\hat{\mathcal{V}}|$ & $\#$ of keys & Max width & Form GCS (s) & Solve (s) & $\delta_{relax}$ $(\%)$ \\
% \midrule
% 1 & 21 & 114 & 2 & 1 & 0.00585 & 0.0408 & 0\\
% 2 & 23 & 160 & 3 & 1 & 0.00777 & 0.0536 & 0\\
% 3 & 74 & 974 & 5 & 2 & 0.0321 & 0.274 & 0\\
% 4 & 78 & 3808 & 5 & 7 & 0.137 & 3.87 & 0.947\\
% 5 & 82 & 5728 & 5 & 10 & 0.214 & 4.53 & 0.363\\
% 6 & 273 & 5786 & 10 & 1 & 0.174 & 1.46 & 0\\
% 7 & 274 & 11480 & 10 & 3 & 0.4 & 25.5 & 0.744\\
% 8 & 284 & 56976 & 10 & 16 & 2.11 & 545 & 0.126\\
% 9 & 1468 & 11720 & 3 & 1 & 0.347 & 3.98 & 0\\
% \bottomrule
% \end{tabular}
% % \caption*{\footnotesize Numerical experiments for various environments using our key-door maze generator.}
% \label{tab:experiments results}
% \end{table*}

In this section, we demonstrate and evaluate the proposed approach to solving a variety of key-door environments. Two benchmarks from \cite{kurtz2023temporal} are shown in Sections V.A. Then we present a method to generate benchmark mazes with keys and doors and present numerical experiments on instances created with this benchmark generator. Code for reproducing these results will be provided in an open-source repository.  All experiments were performed on a laptop with an Apple M2 CPU and 8GB RAM. The underlying convex optimization solver was MOSEK \cite{mosek}, accessed via the GCS components from the Drake \cite{drake} Python bindings. The source code and numerical results are publicly available at \url{https://github.com/TSummersLab/gcspy-precedence-specifications}.

% The augmented GCS construction time and final solving time are shown in Table 1. Since forming augmented GCS is an offline process, it takes partitioned environment as input and computed a GCS as output, we also compare the construction time to the sum of "LTL to DFA" and "Form GCS" time in \cite{kurtz2023temporal}, results are shown in Table 2.
\vspace{-5pt}
\subsection{Key-Door Environments from \cite{kurtz2023temporal}}
\textbf{2-Key Environment.} We first consider the simple key-door environment with two keys and two doors shown in Fig. \ref{fig:simple key door}. The robot must pick up two keys to open two doors in order to reach the target set, which can be written as the STL formula $\varphi= \mathcal{K}_1 \mathcal{R} \neg \mathcal{D}_1 \wedge \mathcal{K}_2 \mathcal{R} \neg \mathcal{D}_2 \wedge \mathcal{F} \, \mathcal{T}$ (which in this case is equivalent to the formula $\neg \mathcal{D}_1 \mathcal{U} \mathcal{K}_1 \wedge \neg \mathcal{D}_2 \mathcal{U} \mathcal{K}_2 \wedge \mathcal{F} \, \mathcal{T}$).

We used Algorithm \ref{partition} to create an exact convex partition and labeled GCS. Next we used Algorithm \ref{Algo1} to construct the augmented GCS, shown in Fig. \ref{fig:augGCS}. Then we used the GCS components in Drake \cite{drake} and the optimization solver Mosek \cite{mosek} to solve a shortest path problem for the augmented GCS. The augmented GCS construction time is \textbf{0.0108} seconds and the solving time is \textbf{0.249} seconds. Compared with an approach that uses general-purpose temporal logic tools \cite{kurtz2023temporal}, our augmented GCS construction is about 80x faster.

\textbf{5-Key Environment.}
We also solved a 5 key-door environment from \cite{kurtz2023temporal}. The STL formula can be written as
\begin{equation}
\begin{split}
    \varphi = &\mathcal{K}_1 \mathcal{R}  \neg \mathcal{D}_1 \wedge \mathcal{K}_2 \mathcal{R}  \neg \mathcal{D}_2 \wedge \mathcal{K}_3 \mathcal{R}  \neg \mathcal{D}_3 \\ &\wedge \mathcal{K}_4 \mathcal{R}  \neg \mathcal{D}_4 \wedge \mathcal{K}_5 \mathcal{R}  \neg \mathcal{D}_5 \wedge \mathcal{F} \, \mathcal{T}.
\end{split}
\end{equation}

The exact convex partitioning from Algorithm \ref{partition} produces a graph with 25 free space cells, 5 key cells, 5 door cells. The augmented GCS constructed by Algorithm \ref{Algo1} contains six layers with nine subgraphs.  The augmented GCS construction time is \textbf{0.0423} seconds and the solving time is \textbf{1.573} seconds. Compared with an approach that uses general-purpose temporal logic tools \cite{kurtz2023temporal}, our augmented GCS construction is more than 50000x faster, and our solve time is about 3.5x faster.

\vspace{-5pt}

\subsection{Key-Door Maze Benchmark Generator}
% \textbf{Benchmark Generator.}
To quickly generate multiple instances of maps with doors and keys to test the algorithm, a maze benchmark generator was created. The generator creates key-door mazes in three steps. First, a \textit{perfect maze} is generated using Eller's algorithm \cite{buck2015mazes}. A perfect maze has a tree structure, where there is a unique path between any two cells in the maze. 
% An example of a perfect maze and an imperfect maze can be seen in Fig. \ref{fig:perfect_vs_imperfect_maze}. 
Second, the start and target cells are assigned randomly, and doors and keys are placed based on a series of breadth-first searches (described in more detail below). All mazes are feasible by construction and have the same number of keys and doors. Third, optionally, walls can be added and/or removed randomly to adjust the width of the associated augmented graph and potentially create infeasibility. This step makes the mazes imperfect since the removal of walls creates branching paths and loops through the environment, making them more challenging to solve. 
% Example mazes are shown in Fig. \ref{fig:narrow_vs_wide_maze}.

Finally, a graph of convex sets is created to represent connectivity among free space, key, and door cells.

% \begin{figure}[htbp]
%   \centering
%   \begin{subfigure}[b]{0.45\linewidth}
%     \centering
%     \includesvg[width=\linewidth]{figures/perfect_maze.svg}
%     \caption{}
%     \label{fig:perfect_maze}
%   \end{subfigure}
%   \hfill
%   \begin{subfigure}[b]{0.45\linewidth}
%     \centering
%     \includesvg[width=\linewidth]{figures/imperfect_maze.svg}
%     \caption{}
%     \label{fig:imperfect_maze}
%   \end{subfigure}
%   \caption{Dark gray cells represent walls and light blue cells represent walkable cells. Maze types: (a) an example of a perfect maze, (b) an example of an imperfect maze.}
%   \label{fig:perfect_vs_imperfect_maze}
% \end{figure}

\textbf{Maze Generator.}
Eller's algorithm rapidly generates a perfect maze one row at a time \cite{buck2015mazes}. It is faster and more memory efficient than alternative maze generation algorithms such as recursive backtracking because it works with at most two rows at once. The number of rows and columns is required to be odd. For any natural numbers $r$ and $c$, a maze is created with $2r + 1$ rows and $2c + 1$ columns. The starting cell is then randomly selected to be in the center or in one of the corners. The target cell is selected to be the cell farthest from the starting cell based on a breadth-first search.

\textbf{Key-Door Batches.}
To create a variety of augmented GCS structures, doors and keys are placed in \textit{batches}. A batch is an integer composition of the total number of key-door pairs and represents how many keys will be accessible either at the start or after a certain number of doors has been unlocked. For example, with 10 key-door pairs, the batch $(1,1,2,1,1,1,1,2)$ creates a narrow maze, with only 1 or 2 keys accessible after opening any door, and the batch $(4,6)$ creates a wide maze, with 4 keys accessible from the start and 6 keys accessible after opening the first 4 doors, respectively.
% These batches are determined by the maximum number of doors. The maximum number of doors serves as an upper bound and does not guarantee that this many doors will be placed. The maximum number of doors can either be set to a natural number or will be calculated based on the size of the maze. An integer composition is created from the maximum number of doors.  The number of batches are randomly created, but can also be controlled by setting the parameter \texttt{graph\_type} to \texttt{'n'} or \texttt{'w'} for narrow and wide, respectively. A narrow maze is created with many batches with fewer doors, while a wide maze is created with fewer batches.  
% An example of a narrow maze and  a wide maze can be seen in Fig. \ref{fig:narrow_vs_wide_maze}. 
% There is a higher probability to place the maximum number of doors in the wide configuration compared to the narrow configuration. The doors and keys will be less spread out in the wide configuration compared to the narrow configuration.

% \begin{figure}[htbp]
%   \centering
%   \begin{subfigure}[b]{0.45\linewidth}
%     \centering
%     \includegraphics[width=\linewidth]{figures/narrow_maze.pdf}
%     \caption{}
%     \label{fig:narrow_maze}
%   \end{subfigure}
%   \hfill
%   \begin{subfigure}[b]{0.45\linewidth}
%     \centering
%     \includegraphics[width=\linewidth]{figures/wide_maze.pdf}
%     \caption{}
%     \label{fig:wide_maze}
%   \end{subfigure}
%   \caption{Example mazes: (a) narrow maze with key-door batch (1, 2, 1, 1, 2), (b) wide maze with key-door batch (7).}
%   \label{fig:narrow_vs_wide_maze}
%   \vspace{-5pt}
% \end{figure}

\textbf{Door Placement.}
For each batch the doors are placed first then the keys. The doors are placed along a path that starts at a target cell and ends at the start cell. For the first batch, the goal cell is the same as the target cell and for the rest of the batches the goal cell is the closest key to the start. The doors are placed in hallways, never in intersections or corners. The larger the batch the closer each door will be relative to each other and the smaller the batch the more spread out each door will be. If the path between the last door is too short or the door is too close to the start it will not be placed. This allows room for the keys to be placed. In the case where the number of doors placed is less than the current batch size, the current batch size will be reduced to equal the number of doors.

\textbf{Key Placement.}
After at least one door has been placed, the keys are placed. The first key is placed in the furthest cell from the start. If the current batch is larger than 1, sequential keys are placed in the dead ends and corners near this key. If the current batch is larger than 1 and the number of keys is less than the number of doors, then one door is removed. Another attempt is then made to place the same number of keys as doors. This process continues until the number of keys is the same as the number of doors. For example, if the number of doors placed is 5 but only 3 keys were placed one door is removed and it attempts to place 4 keys. If only 3 keys could be placed then one more door is removed and it again attempts to place 3 keys. This converts the current batch from 5 to 3. In the case where no keys can be placed and one door remains then that door is removed. After one batch is complete the next one begins in a similar process except the new goal cell becomes the closet key to the start. If no keys are placed the remaining batches, if any, are skipped.

\begin{table*}
\centering
\caption{Numerical Experiments with Key-Door Mazes}
\begin{tabular}{>{\centering\arraybackslash}p{0.03\textwidth}
                >{\centering\arraybackslash}p{0.05\textwidth}
                >{\centering\arraybackslash}p{0.05\textwidth}
                >{\centering\arraybackslash}p{0.08\textwidth}
                >{\centering\arraybackslash}p{0.08\textwidth}
                >{\centering\arraybackslash}p{0.15\textwidth}
                >{\centering\arraybackslash}p{0.10\textwidth}
                >{\centering\arraybackslash}p{0.14\textwidth}}
\toprule
ID & $|\mathcal{V}|$ & $|\hat{\mathcal{V}}|$ & $\#$ of keys & Max width & Form GCS (s) & Solve (s) & $\delta_{relax}$ $(\%)$ \\
\midrule
1 & 21 & 114 & 2 & 1 & 0.00585 & 0.0408 & 0\\
2 & 23 & 160 & 3 & 1 & 0.00777 & 0.0536 & 0\\
3 & 74 & 974 & 5 & 2 & 0.0321 & 0.274 & 0\\
4 & 78 & 3808 & 5 & 7 & 0.137 & 3.87 & 0.947\\
5 & 82 & 5728 & 5 & 10 & 0.214 & 4.53 & 0.363\\
6 & 273 & 5786 & 10 & 1 & 0.174 & 1.46 & 0\\
7 & 274 & 11480 & 10 & 3 & 0.4 & 25.5 & 0.744\\
8 & 284 & 56976 & 10 & 16 & 2.11 & 545 & 0.126\\
9 & 1468 & 11720 & 3 & 1 & 0.347 & 3.98 & 0\\
\bottomrule
\end{tabular}
\vspace{-10pt}
% \caption*{\footnotesize Numerical experiments for various environments using our key-door maze generator.}
\label{tab:experiments results}
\end{table*}

\textbf{Removing or Adding Walls.}
Once keys and doors have been placed, the maze can be modified by removing a random set of walls. Candidate walls for removal are those not on the border, next to a door, or a corner. Each candidate wall is removed with a specified probability, which is a parameter in the code. Removing walls has the effect of widening the associated augmented graph by increasing the number of accessible keys at each layer. It may also make some or even all keys unnecessary.
% \begin{figure}[htbp]
%   \centering
%   \includesvg[width=0.75\linewidth]{figures/candidate_walls.svg}
%   \caption{A perfect maze where red dots represent candidate walls.}
%   \label{fig:candidate_walls}
% \end{figure}
Additionally, the maze can be altered and potentially made infeasible by randomly adding a wall in a hallway, which may block access to a door, key, or the goal. 
% If the maze is opened there is a chance that an infeasible maze becomes feasible. An example of an infeasible maze and an infeasible maze made feasible can be seen in Fig. \ref{fig:infeasible_vs_feasible_maze}.

\begin{figure}
  \centering
  \begin{subfigure}[b]{0.45\linewidth}
    \centering
    \includegraphics[width=\linewidth]{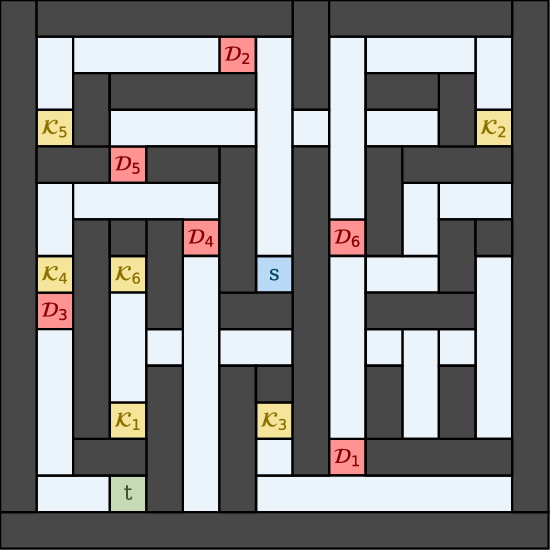}
    \caption{}
    \label{fig:maze_for_graph}
  \end{subfigure}
  \hfill
  \begin{subfigure}[b]{0.45\linewidth}
    \centering
    \includegraphics[width=\linewidth]{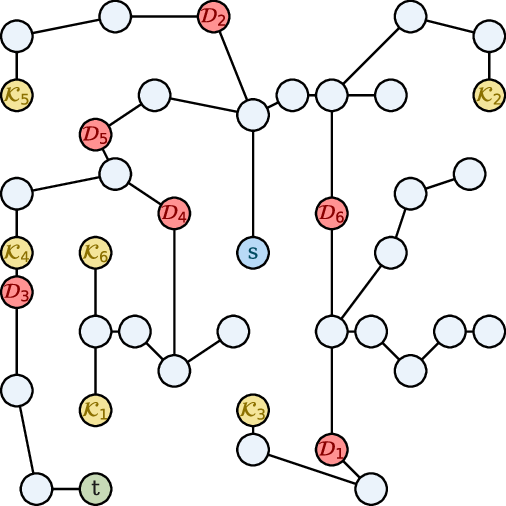}
    \caption{}
    \label{fig:graph_of_maze}
  \end{subfigure}
  \caption{(a) key-door maze, (b) graph of maze.}
  \label{fig:maze_graph_representation}
  \vspace{-5pt}
\end{figure}

% \begin{figure}[htbp]
%   \centering
%   \begin{subfigure}[b]{0.45\linewidth}
%     \centering
%     \includesvg[width=\linewidth]{figures/infeasible_maze.svg}
%     \caption{}
%     \label{fig:infeasible_maze}
%   \end{subfigure}
%   \hfill
%   \begin{subfigure}[b]{0.45\linewidth}
%     \centering
%     \includesvg[width=\linewidth]{figures/feasible_infeasible_maze.svg}
%     \caption{}
%     \label{fig:feasible_infeasible_maze}
%   \end{subfigure}
%   \caption{Infeasible mazes are not solvable such as (a), however, removing walls has the potential to make an infeasible maze feasible such as (b).}
%   \label{fig:infeasible_vs_feasible_maze}
% \end{figure}
\textbf{Graph Creation and Cell Merging.}
% The cells themselves can be squares, rectangles, or jittered squares. These can be selected with the parameter $\texttt{cell\_type}$ which accepts $\texttt{'s'}$, $\texttt{'r'}$, or $\texttt{'j'}$ for squares, rectangles, or jittered squares, respectively. 
Since the cells are arranged in a grid, it is easy to construct a connectivity graph of the free space, key, and doors cells, so we do not use Algorithm \ref{partition} for partitioning. Further, adjacent cells are merged to reduce the number of nodes in the graph. Free space cells are first merged horizontally, then vertically. The vertex representations for the merged free space cells, and for key and door cells are stored.
% If the cells are jittered squares no cells are merged. Fig. \ref{fig:cell_types} shows the same maze with different cell types.
% \begin{figure}[htbp]
%   \centering
%   \begin{subfigure}[b]{0.45\linewidth}
%     \centering
%     \includesvg[width=\linewidth]{figures/square_cell_maze.svg}
%     \caption{}
%     \label{fig:square_cell_maze}
%   \end{subfigure}
%   \hfill
%   \begin{subfigure}[b]{0.45\linewidth}
%     \centering
%     \includesvg[width=\linewidth]{figures/rectangle_cell_maze.svg}
%     \caption{}
%     \label{fig:rectangle_cell_maze}
%   \end{subfigure}
%   \par\bigskip
%   \begin{subfigure}[b]{0.45\linewidth}
%     \centering
%     \includesvg[width=\linewidth]{figures/jitter_cell_maze.svg}
%     \caption{}
%     \label{fig:jitter_cell_maze}
%   \end{subfigure}
%   \caption{Cell types: (a) square cells with simple partition, (b) rectangle cells with simple partition, (c) jittered cells with no partition.}
%   \label{fig:cell_types}
% \end{figure}
After merging, cell vertices are stored with labels $\texttt{'s'}$, $\texttt{'t'}$, $\texttt{'d\#'}$, $\texttt{'k\#'}$, or $\texttt{'c\#'}$ for start, target, door, key, or cell, along with an edge list. Fig. \ref{fig:maze_graph_representation} shows a partitioned and merged maze.

% \subsection{Key-Door Environments from Video Games}
% Demonstrate performance on dungeon environments from The Legend of Zelda series.

% \begin{figure}[htbp]
%   \centering
%   \includegraphics[width=0.80\linewidth]{figures/Zelda Dundeon Room11.drawio.png}
%   \caption{Zelda Dungeon Room 11.}
%   \label{fig:simple}
% \end{figure}

\subsection{Numerical Experiments on Key-Door Mazes}
We evaluate the performance of the proposed approach on a set of key-door mazes created by the our generator. For each maze, we used Algorithm \ref{Algo1} to construct the augmented GCS and then used the GCS components in Drake \cite{drake} and the optimization solver Mosek \cite{mosek} to solve a shortest path problem for the augmented GCS. In particular, we solve a convex relaxation of the exact mixed-integer convex reformulation and obtain a suboptimal solution via a rounding scheme, as described in \cite{marcucci2023motion,marcucci2024shortest}. Table \ref{tab:experiments results} shows the results of nine experiments. For each experiment, we report the number of vertices in the base GCS, $|\mathcal{V}|$; the number of vertices in the augmented GCS, $|\hat{\mathcal{V}}|$; the number of keys; the maximum number of subgraphs in any layer of the augmented GCS, which we call the max width; the augmented GCS construction time; the augmented GCS solving time; and an upper bound on the optimality gap, given by
\[\delta_{relax}=\frac{C_{round}-C_{relax}}{C_{relax}},\]
where $C_{relax}$ is the cost of the relaxation, which lower bounds the optimal value, and $C_{round}$ is the cost of the rounded solution.

\begin{figure}
  \centering
  \includegraphics[width=0.8\linewidth]{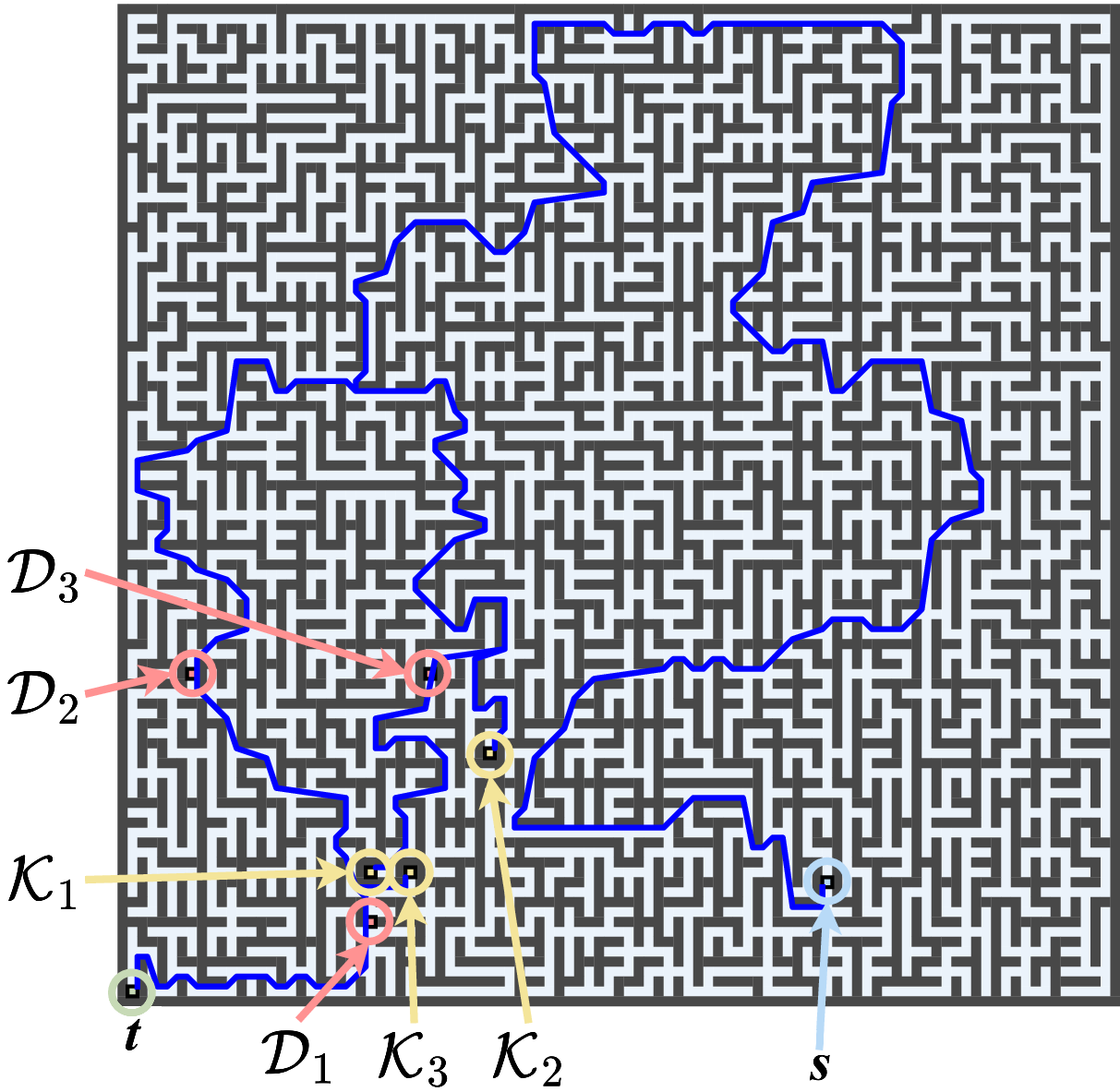}
  \caption{A 99$\times$99 maze with 3 keys and 3 doors.}
  \label{fig:large maze}
  \vspace{-5pt}
\end{figure}

\textbf{Discussion.}
We make several observations based on the results in Table \ref{tab:experiments results}. First, our approach scales to much larger problems than reported in \cite{kurtz2023temporal}. In experiment 6, we solve a maze with 10 keys within 2 seconds. In experiment 9, we solve a large maze with 3 keys within 4 seconds, where the base GCS with 1468 vertices and the augmented GCS has 11720 vertices. The solution is shown in Fig. \ref{fig:large maze}. In all instances the rounded solution is within 1\% of the global optimum, and in most cases is certified as globally optimal.

The max width has a significant effect on the size of the augmented GCS and the solve time. Experiment 8 has 10 keys and a max width of 16, yielding an augmented GCS with 56976 vertices and a solve time of 545 seconds. Further, the instances with non-zero optimality gap are the wider mazes with more possible orders in which to pick up keys. Scaling to environments with many keys and large width will likely require heuristics that trade off computation time with solution quality, which will be explored in future work. 

% \vspace{-10pt}
% \begin{itemize}
%     \item \textbf{Effect of keys.}  
%     For a fixed base GCS size, increasing the number of keys enlarges the augmented GCS.  
%     \emph{Example:} compare \emph{Maze~1} and \emph{Maze~2}, where the higher key count leads to a larger augmented GCS size.

%     \item \textbf{Effect of maximum width.}  
%     With the same base GCS size, a greater maximum width produces a substantial growth in the augmented GCS.  
%     \emph{Example:} the augmented GCS of \emph{Maze~3} through \emph{Maze~5} expands dramatically as the maximum width increases.

%     \item \textbf{Solve time vs.\ construction time.}  
%     The time required to solve the augmented GCS scales more severely than the time required to construct it.  
%     This trend is evident when comparing construction and solve times as the cardinality of the augmented GCS vertices increases.

%     \item \textbf{Optimality gap.}  
%     The optimality gap is more likely to be zero when the augmented GCS is either small in size or relatively narrow.  
%     \emph{Example:} the augmented GCSs of \emph{Mazes~1--3} are small and exhibit zero gaps, while \emph{Mazes~6} and \emph{9} remain narrow and likewise show zero gaps.

%     \item \textbf{Influence of narrow structures on solve time.}  
%     Narrow augmented GCS can significantly reduce solve time.  
%     \emph{Example:} \emph{Maze~6} solves faster than \emph{Maze~5}, and \emph{Maze~9} solves faster than \emph{Maze~7}, despite comparable augmented sizes.
% \end{itemize}

% \section{Conclusions and Outlook}

\bibliographystyle{IEEEtran}
\bibliography{bibliography.bib}

\end{document}